\documentclass[fleqn,twoside]{article}
\usepackage{espcrc2}

\usepackage{graphicx}
\usepackage[figuresright]{rotating}
\usepackage[displaymath]{lineno}

\newcommand{\AmS}{{\protect\the\textfont2
  A\kern-.1667em\lower.5ex\hbox{M}\kern-.125emS}}

\title{The angular resolution of the
Pierre Auger Observatory}

\author{C. Bonifazi%
\address[LPNHE]{LPNHE, Universit\'e Paris VI-VII -- CNRS/IN2P3. 4 place Jussieu, 75252 PARIS CEDEX 05, France}%
\thanks{bonifazi@lpnhe.in2p3.fr}
for the Pierre Auger Collaboration%
\address[]{Pierre Auger Observatory, av. San Mart\'in Norte 304, (5613) Malarg\"ue, Argentina}}

\begin{document}

\begin{abstract}

We discuss the angular resolution obtained for events registered with the surface detector alone and for hybrid events, i.e., those observed simultaneously by both the surface and fluorescence detectors. The angular accuracy of the surface detector is directly extracted from the data itself and on an event by event basis, and is given as a function of the number of stations triggered by the event and of the zenith angle of the shower. We compare the angular resolution of the surface detector obtained from hybrid events with the one obtained from the surface detector alone.  
\vspace{1pc}
\end{abstract}

\maketitle

\section{Introduction}

To fulfill the goal of the Pierre Auger Observatory of determining the origin of the ultra high energy cosmic rays, their arrival direction must be obtained with optimal accuracy. This includes a precise measure of the angular resolution ($AR$), which we define as being the angular radius that contains 68\% of the showers coming from a given point source.

The Pierre Auger Observatory consists of two independent components: the surface detector (SD), which is comprised of 1600 water Cherenkov detectors distributed in a triangular grid with a separation of 1.5~km and covering an area of 3000~km$^2$; and the fluorescence detector (FD), which is formed by 24 fluorescence telescopes located in 4 buildings overviewing the surface detector~\cite{auger}. Hybrid events, i.e., events observed simultaneously by both components, have smaller reconstruction uncertainties than the events observed with only the surface component. On the other hand, the latter have much higher statistics than the former.

The angular resolution for hybrid events is determined from simulations, by computing the angle between the simulated event and its reconstructed direction (section~\ref{arhyb}). The angular resolution for the SD is determined, on an event by event basis, from the zenith ($\theta$) and azimuth ($\phi$) uncertainties obtained from the geometrical reconstruction (section~\ref{arsd}), using the relation~\cite{ICRC2005,ICRC2007}:  
\[
F(\eta) = 1/2~(V[\theta] + \sin^2({\theta})~V[\phi]) 
\]
\noindent
where $\eta$ is the angle between the reconstructed shower direction and the true one, and $V[\theta]$ and $V[\phi]$ are the variances of $\theta$ and $\phi$ respectively. If $\theta$ and $\phi/\sin(\theta)$ have Gaussian distributions with variance $\sigma^2$, then $F(\eta)=\sigma^2$ and $\eta$ has a distribution proportional to $e^{-\eta^2/2\sigma^2}~d(\cos(\eta))~d\phi$. Then, according to our definition, the angular resolution is $AR=1.5~\sqrt{F(\eta)}$.

The arrival direction of a SD event is determined by fitting the arrival time of the first particle in each station according to a shower front model. The accuracy achieved in the arrival direction depends on the clock precision of the detector and on the fluctuations in the  arrival time of the first particle. The timing uncertainty is directly modeled from the data in each station~\cite{ourpaper}~(section~\ref{model}). The model is adjusted using pairs of adjacent stations located in the surface array. Then, it is validated by studying the $\chi^2$ probability distribution for the geometrical reconstruction and by comparing two independent reconstructions of the same event (section~\ref{val}). The angular resolution can therefore be estimated for the SD  reconstruction on an event by event basis (section~\ref{ar}). Using the hybrid events, we are able to extract, for a subset of events, the angular resolution of the surface detector and compare it with the one obtained on an event by event basis (section~\ref{hybrids}).

\section{Angular resolution of hybrid events}\label{arhyb}

The angular resolution for hybrid events is determined from simulations, by computing the angle ($\eta$) between the injected shower axis and the reconstructed one. We simulated 5678 proton showers with the Corsika~6.616 code~\cite{corsika} using QGSJetII-03~\cite{qgsjet2} and FLUKA~\cite{fluka} codes for high and low energy hadronic interaction models respectively. The showers were simulated up to a zenith angle ($\theta$) of 65$^\circ$, with a distribution of $\cos(\theta)\sin(\theta)$, an uniform distribution in the azimuth angle, and for various energies from 0.1 EeV to 10 EeV in steps of 0.25 in logarithmic scale. For each shower we generated uniformly a core position in a slice of 60$^\circ$ in front of one of the fluorescence telescopes with a maximum distance from the telescope increasing with the energy according to the trigger efficiency~\cite{trigger}. Once the hybrid simulation was performed, we reconstructed the simulated events with the same reconstruction chain used for data.

\begin{figure}[t]
\begin{center}
\includegraphics[width=0.45\textwidth]{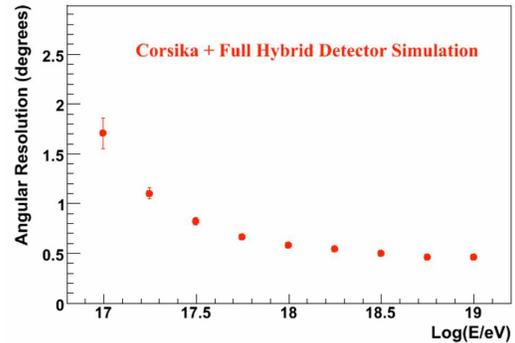}
\end{center}
\vspace{-1cm}
\caption{Angular resolution of hybrid events determined with simulations as a function of the energy. 
\label{fig:arhyb}}
\vspace{-0.5cm}
\end{figure}

In figure~\ref{fig:arhyb} we show the angular resolution, determined as the value where the cumulative distribution function of $\eta(\theta)$ reaches 0.68, as a function of the energy. As can be observed, the angular resolution improves for larger energy showers.  It is worthwhile to remark that this behavior originates from the convolution of the dependence of the angular resolution with the different geometrical parameters. Also, it is important to check these results using real data. For example, using stereo events, which have fluorescence telescopes with triggered pixels in more than one building.

\section{Angular resolution of the surface detector}\label{arsd}

\subsection{Time variance model}\label{model}

The angular accuracy of the surface detector events is driven by the precision in the measurement of the arrival time of the shower front ($T_s$)  
at each station~\cite{ourpaper}. The particle arrival time in the shower front can be
described as a Poisson process over a time interval $T$. The arrival time of the first 
particle ($T_1$) is used as the estimator for the
shower front arrival and its distribution function is given by:
\[  
f(T_1) = \frac{1}{\tau}~e^{-\frac{T_1}{\tau}},
\]
\noindent
with $\tau = T/n$, where $n$ is the number of particles\footnote{To calculate the number of particles $n$, we assumed that the muons are the particles that contribute the most to the time measurement and in average cross the detector with the direction given by the zenith angle of the shower. The number of particles is calculated as the ratio between the total signal in the stations and the track length of the particles.} measured during the time $T$. Then, the variance of $T_1$ is $\tau ^2$, but since we estimate the parameter $T$ from the data itself, it is modified to~\cite{ourpaper}:
\[
V[T_1] = \left( \frac {T}{n}\right )^2~\frac {n-1}{n+1}.
\]
\noindent
The variance of $T_s$ in the SD stations is given by the sum of the detector clock precision ($b^2$) and of the variance of $T_1$. It then becomes: 
\[
V[T_s] = a^2~\left(\frac {2~T_{50}}{n}\right)^2~\frac {n-1}{n+1} + b^2,
\]
\noindent
where $T_{50}$ is the time interval that contains the first 50\% of
the total signal as measured by the photomultiplier FADC (flash analog-to-digital converters) traces. The parameter $a$ is a scale factor, containing all the assumptions considered in the model and the treatment done to the FADC traces. The parameter $b$ should be given by the GPS clock accuracy (about 10 ns) and the FADC trace resolution (25/$\sqrt{12}$~ns), that is $b\simeq 12$~ns. Both $a$ and $b$ are determined from the data. 

A special sub-array of pairs of water Cherenkov detectors has been deployed as a part of the surface array. These are adjacent detectors located $\sim$11~m apart, and therefore are sampling the same region of the shower front. To determine the parameters $a$ and $b$ we used all the events that pass our selection criteria~\cite{T5} from April 2004 to the end of August 2008 with at least one pair in the event. There is total of 46416 events, which are used to fit these two parameters yielding:
\[
\begin{array}{ccrcl}
  a &=& 0.60 \pm 0.01,~~~
  \\
  b &=& \left( 14.1 \pm 0.2 \right) \mbox{\rm ns},
\end{array}
\]
\noindent
with a $\chi ^2/ndof = 0.9992$. More details about the time variance model and the fitting procedure can be found in reference~\cite{ourpaper}.

\subsection{Validation of the time variance model}\label{val}

In this section, we show that the model correctly reproduces the uncertainties of the arrival time of the first particle in the stations. 

\begin{figure}[t]
\begin{center}
\vspace{-2cm}
\includegraphics[width=0.45\textwidth]{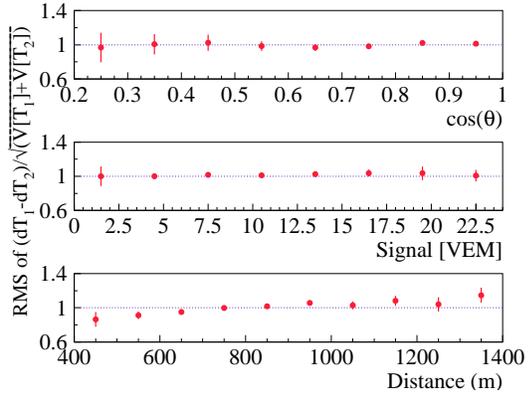}
\vspace{-1cm}
\end{center}
\caption{The RMS of the distribution of $\Delta T /\sqrt{V[\Delta T]}$, as a function of the shower zenith angle (top), the average signal in the adjacent detectors (middle), and their 
distance to the shower core (bottom). 
\label{fig:doublets}}
\vspace{-0.5cm}
\end{figure}

We define the time difference $\Delta T$ = d$T_1 -$~d$T_2$, where d$T_1$ (d$T_2$) is the time residual of the first (second) twin station to the fitted shower front. If the time variance model describes correctly the measurement uncertainties, the distribution of $\Delta T /\sqrt{V[\Delta T]}$, where $V[\Delta T] = V[T_1] + V[T_2]$, should have unit variance. In figure~\ref{fig:doublets} we show the RMS of the distribution of $\Delta T /\sqrt{V[\Delta T]}$ for the adjacent detectors as a function of $\cos(\theta)$ (top), the average signal (middle), and the distance of the paired detectors to the core position (bottom). In all the cases, the RMS is almost constant and close to unity, which shows that the time variance model adequately reproduces the experimental data. It is worthwhile to notice that the time variance model does not explicitly depend on the distance of the station to the shower core. Despite this, the result shown in the bottom panel is satisfactory, with only a small tendency to overestimate the variance for detectors very near to the shower core.

\begin{figure}[t]
\begin{center}
\vspace{-2cm}
\includegraphics[width=0.45\textwidth]{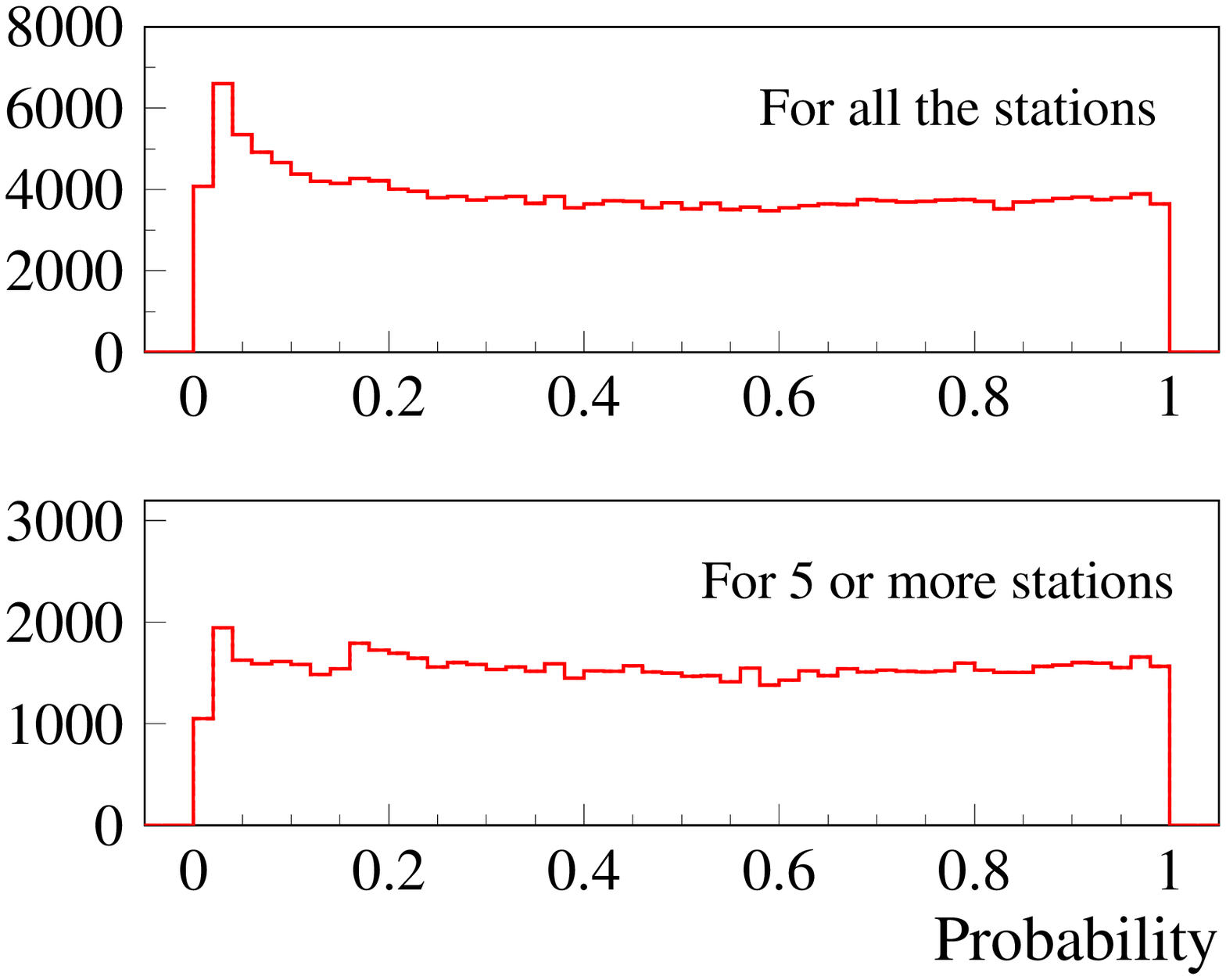}
\vspace{-1.2cm}
\end{center}
\caption{The $\chi ^2$ probability distribution for all events (top) and for 5 or more stations  (bottom). See text for details. 
\label{fig:proba1}}
%
\begin{center}
\vspace{-1.6cm}
\includegraphics[width=0.45\textwidth]{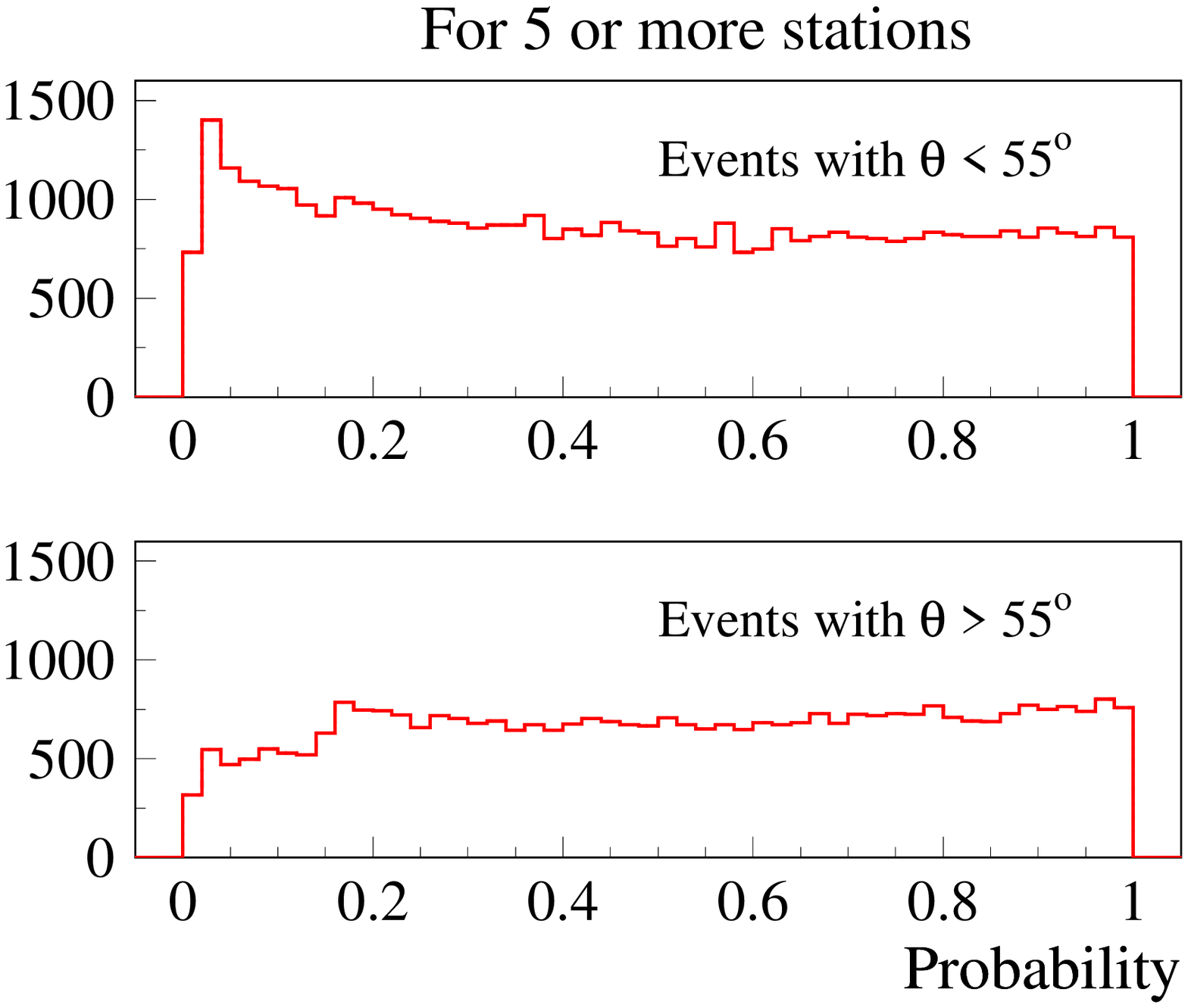}
\vspace{-1.2cm}
\end{center}
\caption{The $\chi ^2$ probability distribution for events with 5 or more stations and with zenith angle smaller (top) and higher (bottom) than 55$^\circ$.
\label{fig:proba2}}
\vspace{-0.5cm}
\end{figure}

We also studied the distribution of the $\chi ^2$ probability of the geometrical reconstruction. In figure~\ref{fig:proba1} (top) we show the distribution for the 308234 events passing our quality cuts~\cite{T5} with 4 or more stations. This distribution is almost flat as it should be in the ideal case, despite the small peak at low values. This small peak could be due, for example, to some stations in the events that have low signal, causing a large $\chi ^2$. But as is seen in figure~\ref{fig:proba1} (bottom), the peak disappears for large multiplicity events (for 5 or more stations). For both cases, we only plot $\chi ^2$ probabilities larger than 1\% to avoid the large peak at zero corresponding to badly reconstructed events ($\sim$ 5\%). Also, we study the $\chi ^2$ probability distribution for two zenith angle ranges, as shown in figure~\ref{fig:proba2}. The $\chi ^2$ probability distribution is flat for both large and small zenith angles, which means that the model works for all angles without compensating one set from the other. This distribution shows that the variance model properly reproduces the uncertainties of the arrival time of the particles in the stations and allows us to determine the angular resolution from the uncertainties in the reconstruction data.

\begin{figure}[t]
\begin{center}
\vspace{-3.8cm}
\hspace{-0.6cm}
\includegraphics[width=0.5\textwidth]{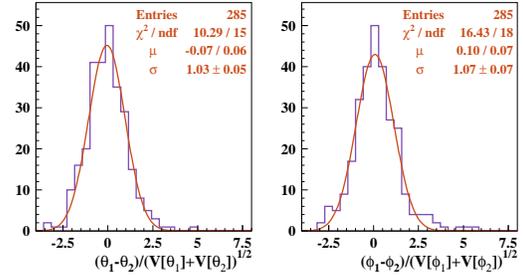}
\vspace{-1.5cm}
\end{center}
\caption{ Distribution of the difference of zenith (left) and azimuth (right) angles from each independent reconstruction divided by the square root of the sum of its variances for the case of 6 or more stations in the event.
\label{fig:check}}
\vspace{-0.5cm}
\end{figure}

We are able to validate independently the time variance model by using the redundant information given by the special sub-array of adjacent detectors. To perform this analysis, we select those events with at least three pairs and with the shower core inside the region defined by these stations, to guarantee a good reconstruction. Then the reconstruction is performed twice, each time using the information of one of the tanks in each pair. This provides two quasi-independent estimates of the arrival direction of the same shower. In the left (right) panel of figure~\ref{fig:check} we show the distribution of the difference of the zenith (azimuth) angles from each reconstruction divided by the square root of the sum of its variances for the case of 6 or more stations which corresponds to events with energy higher than 10 EeV. We fit the distribution with a Gaussian and we obtain a sigma value compatible with 1. Despite of the low statistics, this result shows that the uncertainties in the determination of the arrival time of the first particle in the stations are well represented. The same results are obtained for other multiplicities, except for the 3-fold case, where the sigma is about 0.75 which indicates that we are slightly overestimating the uncertainties. 

\subsection{Angular resolution}\label{ar}

\begin{figure}[t]
\begin{center}
\vspace{-2cm}
\includegraphics[width=0.45\textwidth]{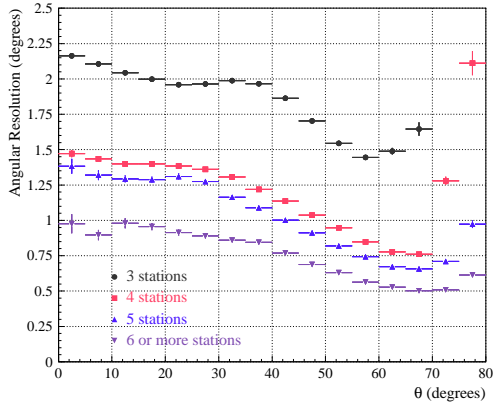}
\vspace{-1.2cm}
\end{center}
\caption{Angular resolution as a function of the zenith angle ($\theta$) and for events with 3 stations (circles), 4 stations (squares), 5 stations (up-triangles), and 6 or more stations (down-triangles).
\label{fig:ar}}
\vspace{-0.5cm}
\end{figure}

Considering the quality of the time variance model for the measurement uncertainties, we can calculate directly the angular resolution on an event by event basis out of the minimization procedure. In figure~\ref{fig:ar}, we show the angular resolution as a function of the zenith angle for various station multiplicities. 

\begin{figure}[t]
\begin{center}
\vspace{-2cm}
\includegraphics[width=0.45\textwidth]{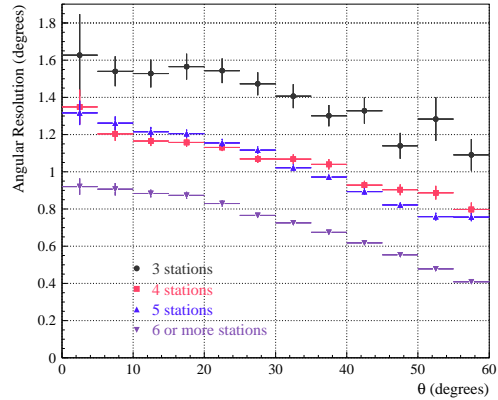}
\vspace{-1.2cm}
\end{center}
\caption{Angular resolution as a function of the zenith angle ($\theta$) for events with an energy above 3 EeV, and for various station multiplicities.
\label{fig:arSat}}
\vspace{-0.5cm}
\end{figure}

The angular resolution is about 2.2$^\circ$ in the worst case of vertical showers with only 3 stations hit. This value improves significantly for 4 and 5 stations. For 6 or more stations, which corresponds to events with energies above 10 EeV, the angular resolution is in all cases better than 1$^\circ$. Above 60$^\circ$, the event multiplicity increases rapidly with zenith angle, and only a few low energy events trigger only 3 stations, thus the accuracy decreases. Also, a $hump$ appears around 40$^\circ$, more visible in the 3-fold 
case. This is due to the contribution of the uncertainties on the core position for events with lower energy (less than few EeV). In figure~\ref{fig:arSat}, we show the angular resolution as a function of the zenith angle for events with an energy above 3 EeV, which is the energy corresponding to a trigger efficiency greater than 99\%~\cite{aperture}. In spite of the degradation of the statistics for 3-fold events, the $hump$ disappears and the angular resolution gets better.  For high multiplicity events, as it is expected, the angular resolution is not affected by the cut in energy. 
We want to remark that all uncertainties quoted in figures~\ref{fig:ar} and~\ref{fig:arSat} are statistical only. We did not, at this stage, investigate possible biases or systematics in the determination of the arrival direction angles.

\section{Angular resolution of the surface detector using hybrid events}\label{hybrids}

Hybrid events that trigger 3 or more stations can be reconstructed using both the hybrid and the surface detector alone modes, giving two independent estimates of the geometry. The comparison of these estimates is therefore an additional independent check of the accuracy on the determination of the arrival direction of the cosmic rays. Therefore, we compute the angle ($\eta$) of those two estimates for different mutiplicities and zenith angle ranges.  Then, the angular resolution of the surface detector can be obtained as:
\[
AR_{SD} = \sqrt{AR_{\eta}^2 - AR_{hyb}^2}
\]
\noindent
where $AR_{\eta}$ is the value of $\eta$ for which the cumulative distribution function of $\eta(\theta)$ reaches 0.68 and $AR_{hyb}$ is the angular resolution of hybrid events obtained from Monte Carlo simulations (shown in figure~\ref{fig:arhyb}). 

\begin{figure}[t]
\begin{center}
\vspace{-2cm}
\includegraphics[width=0.45\textwidth]{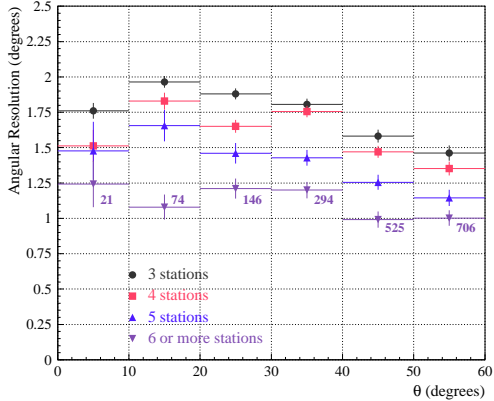}
\vspace{-1.2cm}
\end{center}
\caption{Angular resolution for the surface detector events using the hybrid reconstruction as a reference, as a function of the zenith angle ($\theta$) and for different numbers of stations in the event. See text for more details.
\label{fig:arfromhyb}}
\vspace{-0.5cm}
\end{figure}

In figure~\ref{fig:arfromhyb} we show the angular resolution for the surface detector ($AR_{SD}$) using the hybrid reconstruction as a reference, as a function of the zenith angle ($\theta$) for different numbers of stations in the event. This plot can be directly compared with the one in figure~\ref{fig:ar} despite the differences in the statistics. The values obtained in figure~\ref{fig:arfromhyb} are slightly higher than the ones obtained from the direct determination of the angular resolution on an event by event basis. This could be due to systematic uncertainties in the reconstructions. For the case of 3-fold events, this slight difference is not present and this could be due to the fact that the uncertainties seem to be overestimated (section~\ref{val}).

\section{Summary}\label{sum}

We have determined the angular resolution for the hybrid events using Monte Carlo simulations and we found that for energies above 3 EeV it is about 0.5$^\circ$.  For the surface detector, we determined the angular resolution from the data. This could be done thanks to the time variance model developed to describe the measurement uncertainties of the time arrival of the first particle in the surface stations. Using this model we obtaines an optimal determination of the shower arrival direction and are able to extract the angular resolution of the surface detector on an event by event basis. We found the angular resolution to be better than 2.2$^\circ$ for 3-fold events (E~$<$~4~EeV), about 1.5$^\circ$ for 4-fold and 5-fold events (3~EeV$<$~E~$<$~10~EeV) and better than 1$^\circ$ for higher multiplicity events (E~$>$~10~EeV).


\begin{thebibliography}{9}

\bibitem{auger} J. Abraham {\em et al.} [Pierre Auger Collaboration],
  Nucl. Instr. \& Meth. A 523, (2004)
  
\bibitem{ICRC2005}  C. Bonifazi [Pierre Auger Collaboration], in Proceedings 
of the 29th International Cosmic Ray Conference, Pune, India, vol. 7, pp. 17Ð20, (2005) 

\bibitem{ICRC2007}  M. Ave [Pierre Auger Collaboration], in Proceedings 
of the 30th International Cosmic Ray Conference, M\'erida, M\'exico, (2007); 
arXiv:0709.2125 [astro-ph] 

\bibitem{ourpaper}  C. Bonifazi, A. Letessier-Selvon, E.M. Santos, Astropart. Phys. 28, 523-528 (2008); arXiv:0705.1856 [astro-ph] 

\bibitem{corsika} D. Heck {\em et al.}, {\it ``CORSIKA: A Monte Carlo Code to Simulate Extensive Air Showers''}, Report FZKA 6019, (1998);  D. Heck, T. Pierog, J. Knapp, CORSIKA: an air shower simulation program: http://www-ik.fzk.de/corsika/

\bibitem{qgsjet2} S. Ostapchenko, Nucl. Phys. B (Proc. Suppl.) 151, 143 (2006); 
[arXiv:hep-ph/0412332]

\bibitem{fluka} A. Fasso, A. Ferrari, J. Ranft, P.R. Sala, FLUKA: a multi-particle transport code, CERN-2005-10, 2005.

\bibitem{trigger} A. Ewers {\em et al.} [Pierre Auger Collaboration], in Proceedings 
of the 29th International Cosmic Ray Conference, Pune, India, vol. 7, pp. 115-118, (2005)

\bibitem{T5}  D. Allard  {\em et al.} [Pierre Auger Collaboration], in Proceedings 
of the 29th International Cosmic Ray Conference, Pune, India, vol. 7, pp. 287-290, (2005) 

\bibitem{aperture} D. Allard  {\em et al.} [Pierre Auger Collaboration], in Proceedings 
of the 29th International Cosmic Ray Conference, Pune, India, vol. 7, pp. 71-74 , (2005); [arXiv:astro-ph/0511104]
 
\end{thebibliography}
\end{document}